\documentclass[twocolumn,twoside,slac_two]{revtex4}
\usepackage{graphicx}
\usepackage{fancyhdr}
\newcommand{\etal}{\textit{et al.}}

\begin{document}

\title{Weak Charm Decays with Lattice QCD}

%

\author{A.~X.~El-Khadra}
\affiliation{Department of Physics, University of Illinois, Urbana, IL 61801,
USA}

\begin{abstract}
In this paper I review the status of lattice QCD calculations of
$D$ and $D_s$ meson decay constants and of $D$ meson semileptonic decay
form factors. I restrict my discussion to results obtained from
simulations with $n_f = 2+1$ sea quarks.

\end{abstract}

\maketitle

\thispagestyle{fancy}


\section{Introduction and Motivation}

Lattice QCD is the only systematically improvable calculational tool
we have for quantitatively understanding nonperturbative QCD effects.
Accurate theoretical calculations of nonperturbative QCD effects
are essential for the experimental flavor physics program. One set of
goals of the experimental program are accurate determinations of the CKM
matrix elements. This is illustrated for the weak decay process
$D \rightarrow K l\nu$. The experimentally measured (differential)
decay rate can be written as
\begin{equation}
 \frac{d\Gamma}{d q^2} = {\rm (known)} |V_{cs}|^2 f^2_+(q^2)
\end{equation} \label{exp}
where $f_+(q^2)$ is one of the hadronic form factors which parameterize
the hadronic matrix element for this process,
$\langle K|V_{\mu}|D\rangle$. Hence, to determine $|V_{cs}|$ from
experimental measurements, we need a theoretical calculation of the
form factor with matching precision.
Another set of goals is to
constrain beyond the standard model theories and to search for
new physics signals. This effort complements the experiments at the
high energy frontier. Accurate theoretical calculations are again
essential.

Since lattice QCD calculations are complicated and time consuming, a
third important goal specifically of the charm physics experiments is
to test lattice QCD methods. For example, we can use
Eq.~\ref{exp} to determine the form factors from experimental
measurements after taking $|V_{cs}|$ from other sources. These
tests are important to establish lattice methods for the $B$ meson
system, where the CKM matrix elements are less well known, and where
input from lattice QCD is essential.
The leptonic and semileptonic $D$ meson decays discussed in this
talk are ideal for this. They are not expected to be sensitive to new
physics, and the corresponding hadronic matrix elements are
straightforward to calculate. Once established, lattice QCD together
with the experimental measurements can then be used to improve the
determinations of the CKM angles $V_{cd}$ and $V_{cs}$. All of these
goals require accurate measurements and calculations.

\subsection{Introduction to Lattice QCD}

In lattice field theory, the space-time continuum is replaced by a
discrete lattice. (For reviews of lattice QCD see Ref.~\cite{latrev}.)
This implies that derivatives are replaced by discrete differences,
which in turn introduces discretisation errors into physical quantities.
These errors generally vanish with a positive power of the lattice
spacing ($a$).

Nonperturbative calculations in lattice QCD can be performed using
Monte Carlo methods. Lattice artifacts can be removed by reducing the
lattice spacing used in numerical calculations. However, the
computational cost increases  as $~1/a^7$ (keeping the other parameters
fixed). Alternatively, one can reduce discretisation errors by
adding higher-dimensional operators to the action.
This is called improvement. With improved actions the computational
effort needed to perform reliable lattice QCD calculations can
potentially be significantly reduced. This idea is behind much of the
important progress made in lattice QCD calculations in recent years and
has been an increasing part of research in lattice field theory.

The main obstacle for obtaining quantitative results (at
the few percent level) from numerical simulations of lattice QCD has
always been the computational effort associated with the proper
inclusion of sea quark effects. Several years ago, substantial
progress was made on this problem, in large part due to the
development of an improved staggered fermion action
\cite{asqtad}. For the first time, computationally efficient lattice
simulations with realistic sea quark effects have become possible.

\subsection{Light Quark Methods} \label{lqm}

The simplest lattice quark action replaces the covariant
derivative in the continuum action by a discrete difference
operator. This so-called naive action suffers from the doubling
problem. For every continuum quark flavor, it has 15 additional
unphysical flavors, called tastes. The staggered quark action
combines four of these tastes into one Dirac field, by staggering
the quark fields on a hypercube. This leaves four unphysical flavors
(tastes). This action suffers from large $O(a^2)$ lattice spacing
artifacts due to taste changing interactions. The Asqtad action
is an improved staggered action where all tree-level discretisation
errors are removed \cite{asqtad}. Its leading lattice spacing
errors are therefore of $O(\alpha_s a^2)$ and greatly reduced
compared to the original staggered action.
The Asqtad action is the computationally most efficient light quark
action available. However, in order to use it for the sea
quarks in numerical simulations, the unphysical flavors must
be removed. The sea quarks are present in the fermion determinant of the
path integral. To simulate two degenerate light (up and down) sea quarks,
the MILC collaboration simply takes the square root of the
light quark fermion determinant. For the strange sea quark, they take
the fourth root of the determinant. This procedure is still somewhat
controversial, but there is a growing body of evidence that its
effects are controllable and disappear in the continuum
limit \cite{ask_lat07}.  The Asqtad action with the square root trick
has been extensively tested in numerical simulations, most prominently
in Ref.~\cite{prl}.

The HISQ (highly improved staggered quark) action is another
version of an improved staggered action \cite{hisq}. Like the
Asqtad action, it removes all
tree-level $O(a^2)$ errors. The $O(\alpha_s a^2)$ errors in the
Asqtad action are rather large, due to taste changing
interactions which appear at one-loop order. The HISQ action
reduces the $O(\alpha_s a^2)$ taste-changing effects by roughly a factor
of three over the Asqtad action. The HISQ action has not yet been
used to generate $n_f = 2+1$ sea quark ensembles. Its computational
cost is expected to be about a factor of two larger than the Asqtad
action.

Other light quark methods include the Wilson action and its improvements
\cite{wilson},
Domain Wall Fermions \cite{kaplan} and Overlap fermions \cite{nn}, with
increasing computational cost. The Wilson action solves the doubling
problem by adding a dimension five operator which breaks chiral symmetry. 
Domain Wall Fermions
solve the doubling problem by adding a fifth dimension, while keeping
chiral symmetry almost exact. Overlap fermions have exact chiral symmetry,
but a complicated operator structure.

\subsection{Heavy Quark Methods}

On the lattice, heavy quarks with $am_Q$ large, are best treated
within an effective field theory framework (NRQCD or HQET). One can
start with an effective field theory, and discretise it as in
Ref.~\cite{lepage}, for example. Alternatively, one can start
with a relativistic lattice action and analyze its mass dependent
discretisation errors using effective field theories. The charm
quark it too light for a straightforward implementation of the 
former approach, so we will focus on the latter.

The Fermilab approach \cite{kkm} starts with the improved relativistic
Wilson action \cite{wilson} and the observation that the Wilson action
has the same heavy quark limit as QCD. With a simple prescription, the
Wilson action can be used for heavy quarks without errors that grow
with the heavy quark mass, $(am_Q)^n$. This approach can be used for both
charm and and beauty quarks. With the improved Wilson action, the leading
discretisation errors are $O(\alpha_s \Lambda/m_Q)$ and $O(\Lambda/m_Q)^2$.

The HISQ action is so highly improved that it can be used for charm
quarks with an additional tuning of a parameter in the action, provided
that the lattice spacing is small enough \cite{hisq}.  The leading mass 
dependent discretisation errors are formally of order $O(\alpha_s (am_c)^2)$
and $O(am_c)^4$.

\subsection{Systematic Errors}

The most important sources of systematic error in lattice QCD
calculations are sea quark effects; using unphysically large masses
for the up and down quarks; discretisation effects; finite volume
effects; and renormalisation effects.

In order to be phenomenologically relevant, a lattice QCD
calculation must use gauge configurations that include the effects
of three light sea quarks. Since the masses of the up and down quarks
are generally taken to be degenerate, this is also referred to
as the $n_f = 2+1$ case.

Until roughly five years ago, almost all lattice
QCD calculations used ensembles generated either in the quenched
approximation or with an incorrect number of sea quarks
(generally, $n_f = 2$) because of the computational cost associated with
including sea quarks in the simulations. The quenched approximation
omits sea quark effects entirely, at the cost of adding a systematic
error in the range of $10-30\%$ for physical quantities involving
stable hadrons \cite{prl}. This error must be determined on a case by
case basis. Simulations with an incorrect number of sea quarks carry
a similar systematic error, which is hard to estimate {\it a priori}.

The computational cost increases with decreasing sea quark
mass as $m_l^{-2.5}$. All simulations to date use masses for the light
sea quarks which are larger than the physical up and down quark masses.
(Note, the strange quark mass is large enough to be simulated at its
physical value.)
We can use chiral perturbation theory (ChPT) to guide the extrapolations
from the light sea quark masses used in the simulations to the
physical masses. ChPT is an effective theory of QCD, which can be
applied to (lattice QCD calculations involving) pions and kaons. It
can be combined with heavy quark effective theory and be applied
to heavy-light systems, such as $D$ and $B$ mesons. Furthermore,
it can be extended to include the leading light quark discretisation
errors. Indeed, this has been done for the taste changing interactions
of the Asqtad action and is called staggered ChPT (SChPT) \cite{cb}.
The ChPT extrapolations are a significant but controllable source of
systematic error. In order to keep this error at the few percent level
or less, one needs to include simulations with a range of light sea
masses, keeping $m_l < m_s/2$. The lattice QCD calculations
described here include light sea quarks with masses in the range
$m_l = 1/10\, m_s$ -- $1/2 \,m_s$. Hence, the lightest light sea quark
masses are only a factor of $2-3$ away from their physical value.

As described in the previous sections, the lattice actions give
rise to discretisation errors. They can usually be estimated
{\it a priori} using power counting arguments. However, even with improved
actions, it is important to study and possibly remove these errors
by repeating the calculation at several lattice spacings.

\subsection{Simulation Parameters}

\begin{figure}[htb]
\centering
\includegraphics[clip=true,width=0.40\textwidth,angle=270]{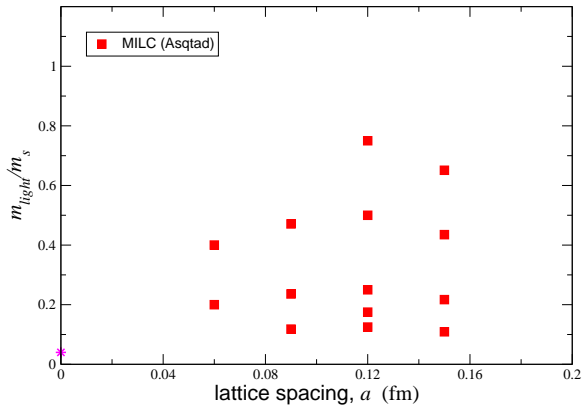}
\caption{Simulation parameters for the MILC ensembles with $n_f = 2+1$.
showing $m_l/m_s$ vs.~lattice spacing $a$ (red squares). The physical
point is at $m_l/m_s = 1/25$ (pink burst). } \label{mva}
\end{figure}

The simulation parameters of the $n_f = 2+1$ sea quark ensembles
generated by the MILC collaboration using the Asqtad action (with the
square root trick) are listed below
and are shown in Figure~\ref{mva}. Each ensemble contains between
450--800 configurations. The ensembles contain one sea quark with
a mass near the strange quark mass, $m_s$, and two degenerate
light sea quarks with masses, $m_l$.
\begin{itemize}
\item $a = 0.15$ fm; $m_l = 0.1 \,m_s$, $0.2 \,m_s$, $0.4\, m_s$, $0.6\, m_s$.
\item $a = 0.12$ fm; $m_l = 0.125 \,m_s$, $0.25\, m_s$, $0.5\, m_s$, 
$0.75\, m_s$.
\item $a = 0.09$ fm; $m_l = 0.1\, m_s$, $0.2\, m_s$, $0.4 \,m_s$.
\end{itemize}

\section{Semileptonic $D$ Meson Form Factors}

The semileptonic decays $D \rightarrow K (\pi) l \nu$ are
mediated by weak vector currents. The hadronic matrix elements
for semileptonic decays are parameterized in terms of form
factors. In our case there are two form factors, conventionally
$f_+(q^2)$ and $f_0(q^2)$. The form factors are functions of the
virtual $W$ boson momentum transfer, $q^2$, or, equivalently, the
recoil momentum of the daughter meson. This introduces additional
lattice spacing errors:
\begin{equation}
\langle K | V_{\mu} | D \rangle^{\rm lat} =
 \langle K | V_{\mu} | D \rangle^{\rm cont} + O(ap_K)^n
\end{equation}
Hence, discretisation errors are smallest, when $p_K$ is
small and $q^2 \approx q^2_{\rm max} = (m_D - m_K)^2$.

The finite lattice volume provides an infrared cut-off, and
therefore a minimum value for finite momentum,
$p_{\rm min} = \frac{2 \pi}{L}$. Lattice three-momenta can be written
in terms of $p_{\min}$ as $\vec{p} = p_{\rm min} (n_x, n_y, n_z)$,
where $n_x,n_y,n_z$ are integers. For example, for $a = 0.1$ fm,
$L = 20$, $p_{\rm min} = 620$ MeV.

To date, the only lattice results for semileptonic $D$ meson form
factors with $n_f = 2+1$ are from the Fermilab Lattice and MILC
collaborations \cite{slprl}. They use the MILC $a = 0.12$ fm lattices
with light sea quark masses in the range $m_l = 1/8 m_s$ -- $3/4 m_s$,
the Asqtad action for the light valence quarks and the Fermilab
action for the charm quark. Staggered chiral perturbation theory
is used to extrapolate to the physical light quark masses and to remove
the leading discretisation errors due to taste violations.

\begin{figure}
\centering
\includegraphics[width=0.40\textwidth]{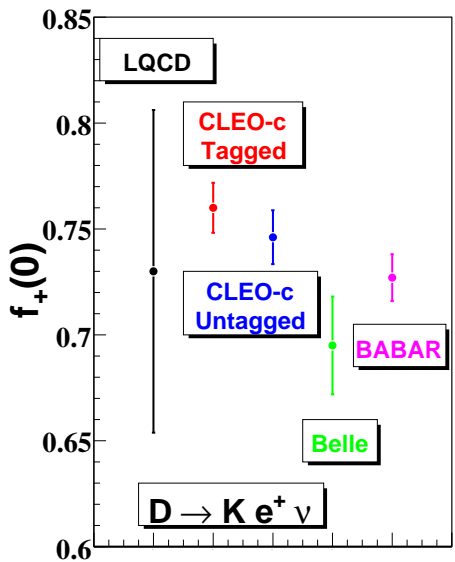}
\caption{$f^K_+(0)$ in comparison from Ref.~\cite{pavlunin}} \label{f0}
\end{figure}

Figure~\ref{f0} shows a comparison of the lattice QCD result
for the normalization $f^K_+(0)$ for $D\rightarrow K l \nu$ with
experimental determinations (where $V_{cs}$ is taken from other
sources). The results are in very good agreement; however, the
lattice QCD result has much larger errors than the experimental
determinations. The comparison between lattice theory and experiment
for $f^{\pi}_+(0)$ is similar \cite{pavlunin}.

\begin{figure}[htb]
\centering
\includegraphics[clip=true,width=0.4\textwidth,angle=270]{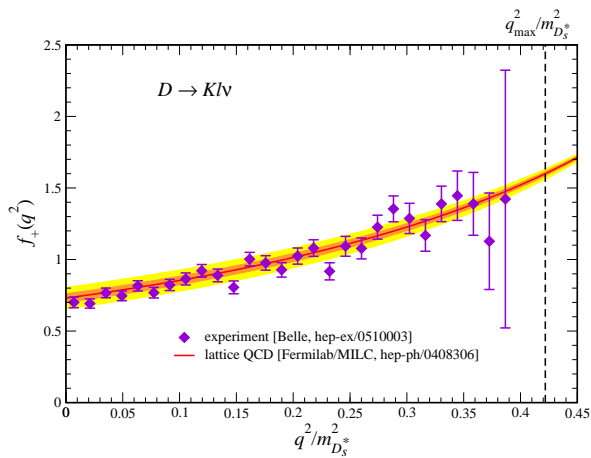}
\caption{The shape of $f^K_+(q^2)$ in comparison. The lattice
prediction for the shape is from Ref.~\cite{ask}.} \label{shape}
\end{figure}

\begin{figure}[htb]
\centering
\includegraphics[clip=true,width=0.4\textwidth,angle=270]{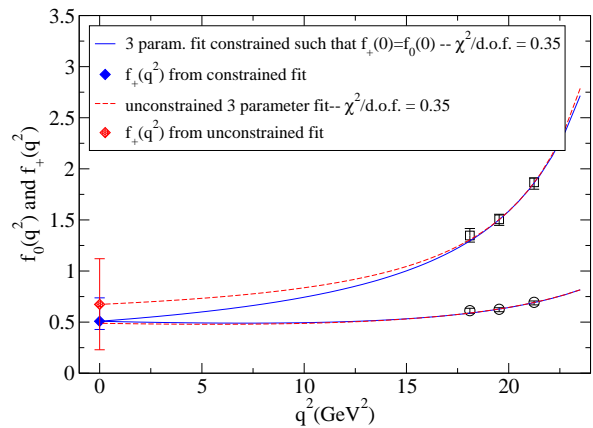}
\caption{The form factors for $B \rightarrow \pi l \nu$ vs.~$q^2$
from fitting to the $z$-expansion. The preliminary results are
from Ref.~\cite{rvdw06}.} \label{btopi}
\end{figure}

The shape of the form factor can also be determined in lattice QCD.
However, in Ref.~\cite{slprl} the form factors were
calculated at a few values of recoil momentum. Then the BK
\cite{bk}
parameterisation was used to determine the $q^2$ dependence of
the form factors. Since the errors increase with recoil momentum,
the shape of the form factors is fixed mainly from the form
factors near $q^2_{\rm max}$ and from using the BK parameterisation
\cite{ask}. The lattice QCD result appeared before the new
measurements by the FOCUS \cite{focus} and Belle \cite{belle} 
collaborations were
announced, so it is one of a very few lattice QCD
{\em predictions}.
Figure~\ref{shape} \cite{ask} shows a comparison of the lattice
prediction for the $q^2$ dependence with experimental data from
the Belle collaboration \cite{belle}. The agreement is excellent.
However, a quantitative comparison between the BK shape
parameter determined from experiment and lattice theory is difficult
to interpret, as eloquently argued by Richard Hill~\cite{hill}.
A model independent parameterisation of the shape based on the
$z$-expansion would avoid this difficulty \cite{arnesen}. The
$z$-expansion is being used by the Fermilab Lattice and MILC
(FNAL/MILC) collaboration to parameterize the $q^2$ dependence
of the form factors for $B \rightarrow \pi l \nu$ \cite{rvdw06}.
This works
quite well, as shown in Figure~\ref{btopi}. Any new lattice
analysis of semileptonic $D$ meson decay form factors will
(should) adopt the $z$-expansion to determine the shape.

A number of additional improvements are possible in future
calculations. Twisted boundary conditions can be used to adjust
the lattice momenta to arbitrarily small values \cite{sach}, which
would improve the shape determination. Double ratio methods
similar to what has been developed for lattice studies of
$B \rightarrow D l \nu$ \cite{btd} can
be adapted to semileptonic $D$ meson decays \cite{haas}.
This may lead to reduced statistical errors as well as improvements
of some of the systematic errors.

\section{Leptonic Decay Constants $f_D$ and $f_{D_s}$}

Charm leptonic decays provide another important test of lattice
QCD. The lattice methods for calculating decay constants in the
charm and beauty meson systems are the same. Indeed, with the
Fermilab approach one uses the same heavy quark action in
both systems and the heavy quark discretisation errors are
expected to be larger for $D$ mesons than for $B$ mesons.

There are now results from two groups (FNAL/MILC and HPQCD).
Both use the MILC ensembles at $a = 0.09$ fm, $0.12$ fm,
$0.15$ fm.

The first FNAL/MILC results came out in 2005 \cite{fdprl},
just days before CLEO-c announced its first precise
determination of $f_{D^+}$ \cite{fdcleo}; the two results
were in good agreement.

The HPQCD collaboration announced their results for decay
constants with much reduced errors this summer \cite{hpqcd} and
FNAL/MILC presented updated results at the Lattice 2007
conference \cite{fdnew}, also with reduced errors. The new
FNAL/MILC analysis was done ``blind'', where an overall unknown
offset was added to the lattice data. The final results were
unblinded shortly before they were presented at the Lattice 2007
conference, making this the first (intentionally) blind lattice 
analysis.
Table~\ref{fdcomp} compares the main features of the two
calculations. More details about the HPQCD and FNAL/MILC calculations,
including discussion of the error analysis and plots of chiral
and lattice spacing extrapolations can be found in Refs.~\cite{ef07}
and \cite{fdnew} respectively.
\begin{table*}[ht]
\begin{center}
\caption{Comparison of the main features of the HPQCD and Fermilab
Lattice/MILC calculations.}
\begin{tabular}{|l|l|l|l|}
\hline \multicolumn{2}{|l|}{FNAL/MILC} & \multicolumn{2}{l|}{HPQCD} \\
\hline
\multicolumn{2}{|l|}{Fermilab action for charm quark} &
\multicolumn{2}{l|}{HISQ action for charm and light valence} \\
\multicolumn{2}{|l|}{Asqtad action for strange and light valence} &
\multicolumn{2}{l|}{\ quarks} \\
\hline
 $a$ (fm) & $m_l/m_s$ sea quark & $a$ (fm) & $m_l/m_s$ sea quark \\ \hline
0.09 & 1/10, 1/5, 2/5 & 0.09 & 1/5, 2/5 \\
0.12 & 1/8, 1/4, 1/2, 3/4 & 0.12 & 1/8, 1/4, 1/2 \\
0.15 & 1/10, 1/5, 2/5, 3/5 & 0.15 & 1/5, 2/5 \\
\hline
\multicolumn{2}{|l|}{$8-12$ light valence quark masses per ensemble} &
\multicolumn{2}{l|}{1 valence quark mass/ensemble, $m_{\rm valence} = m_{\rm sea}$} \\ \hline
\multicolumn{2}{|l|}{Partial nonperturbative renormalisation} &
\multicolumn{2}{l|}{Nonperturbative renormalisation from PCAC} \\
\multicolumn{2}{|l|}{Staggered ChPT fits to all valence} &
\multicolumn{2}{l|}{Continuum ChPT $+O(a^2)$ terms fit to all} \\
\multicolumn{2}{|l|}{\ and sea quark ensembles together} &
\multicolumn{2}{l|}{\ ensembles together} \\
\multicolumn{2}{|l|}{Blind analysis for Lattice 2007} &
\multicolumn{2}{l|}{ } \\
\hline
\end{tabular}
\label{fdcomp}
\end{center}
\end{table*}
The FNAL/MILC analysis includes more lattice ensembles, more
valence quark masses per ensemble, and uses staggered chiral
perturbation theory (Staggered ChPT) to remove the leading light
quark discretisation errors.
The HPQCD collaboration considers only the case $m_q = m_l$,
where $m_q$ denotes the light valence quark mass and $m_l$
denotes the light sea quark mass. They use continuum ChPT with
generic $O(a^2)$ terms added in the chiral fits.

The main difference between the two calculations is the
valence quark actions.
The HPQCD collaboration uses the HISQ action for all (charm,
strange and light) valence quarks, whereas the FNAL/MILC
collaboration uses the Fermilab action for the charm quarks and
the Asqtad action for the strange and light valence quarks.
Since the HISQ action is more improved than the Fermilab
action, the HPQCD result has much smaller heavy quark
discretisation errors. This is the main reason for the difference
in the total errors between the two results.

Table~\ref{error} compares the error budgets for the 2005 FNAL/MILC
calculation with the FNAL/MILC Lattice 2007 one. The error
reduction is mainly due to including three MILC ensembles at
$a = 0.09$ fm (and 8-12 different valence masses). This reduces the heavy
quark and light quark discretisation errors, and better constrains
the staggered ChPT.
\begin{table}[h]
\begin{center}
\caption{ Comparison of the error budget of the 2005 FNAL/MILC
 results with the Lattice 2007 results. Numbers are given in percent.}
\begin{tabular}{|l|l|l|l|l|}
\hline  & \multicolumn{2}{c|}{PRL 2005 \cite{fdprl}} &
\multicolumn{2}{c|}{Lattice 2007 \cite{fdnew}} \\
\hline
source & $f_{D^+}$ & $f_{D_s}/f_{D^+}$ & $f_{D^+}$ & $f_{D_s}/f_{D^+}$ \\
\hline
statistics & 1.5 & 0.5 & 3.8 & 1.0 \\
\hline
HQ discretisation & 4.2 & 0.5 & 2.7 & 0.3 \\
light quark + Chiral fits & 6.3 & 5 & 2.7 & 1.8 \\
inputs ($a$, $m_c$, $m_s$) & 2.8 & 0.6 & 3.4 & 0.5 \\
higher order PT & 1.3 & 1.3 & 0.3 & -\\
\multicolumn{5}{|l|}{ $+$ other small sources (finite volume, $\ldots$)} \\
\hline
total systematic & 8.5 & 5.4 & 5.3 & 2.0 \\ \hline
\end{tabular}
\label{error}
\end{center}
\end{table}

\begin{figure}[htb]
\centering
\includegraphics[clip=true,width=0.40\textwidth,angle=270]{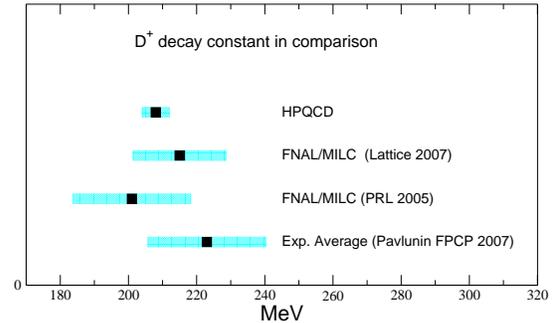}
\caption{Comparison of lattice QCD results for $f_{D^+}$ with experiment.} 
\label{fd}
\end{figure}

\begin{figure}[htb]
\centering
\includegraphics[clip=true,width=0.40\textwidth,angle=270]{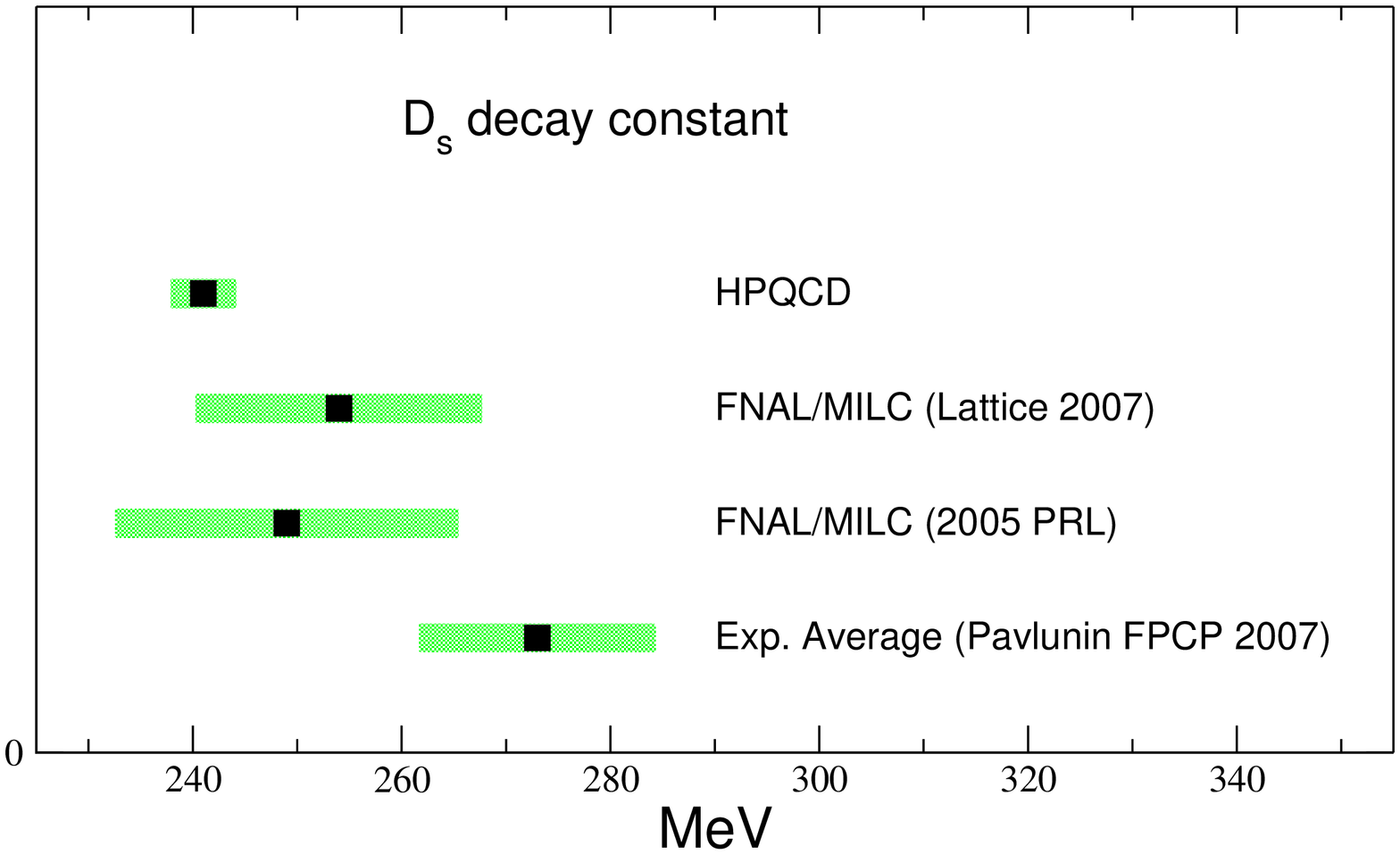}
\caption{Comparison of lattice QCD results for $f_{D_s}$ with 
experiment.} \label{fds}
\end{figure}

\begin{figure}[htb]
\centering
\includegraphics[width=0.40\textwidth,angle=270]{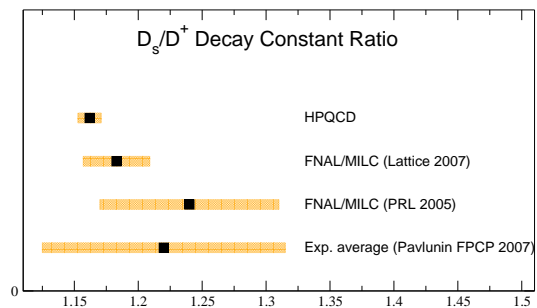}
\caption{Comparison of lattice QCD results for $f_{D_s}/f_{D^+}$ with 
experiment.} \label{fdsfd}
\end{figure}

Figures~\ref{fd}, \ref{fds}, and \ref{fdsfd} compare the lattice
results for $f_{D^+}$, $f_{D_s}$, and
$f_{D_s}/f_{D^+}$, respectively, to the corresponding experimental
averages.  The experimental averages are from
Ref.~\cite{pavlunin}. The new CLEO-c result $f_{D_s} = 275 \pm 10 \pm 5$
presented at this conference by Steven Blusk \cite{blusk}, is very
similar to Ref.~\cite{pavlunin}.


The FNAL/MILC results agree with the experimental averages
at the one sigma level. The HPQCD results agree very well with the
FNAL/MILC results. There is a hint of disagreement between
the HPQCD result for $f_{D_s}$ and the experimental average
at the two sigma level. However, the experimental determinations of
the decay constants must assume a value for the CKM angle $V_{cs}$
from other sources. We are approaching a level of precision, where
tests of lattice QCD should be performed on CKM free
quantities such as the ratio of semileptonic to leptonic decay
rates suggested in Ref.~\cite{cleoc}.

\section{Conclusions and Outlook}

\begin{figure}[h]
\centering
\includegraphics[clip=true,width=0.40\textwidth,angle=270]{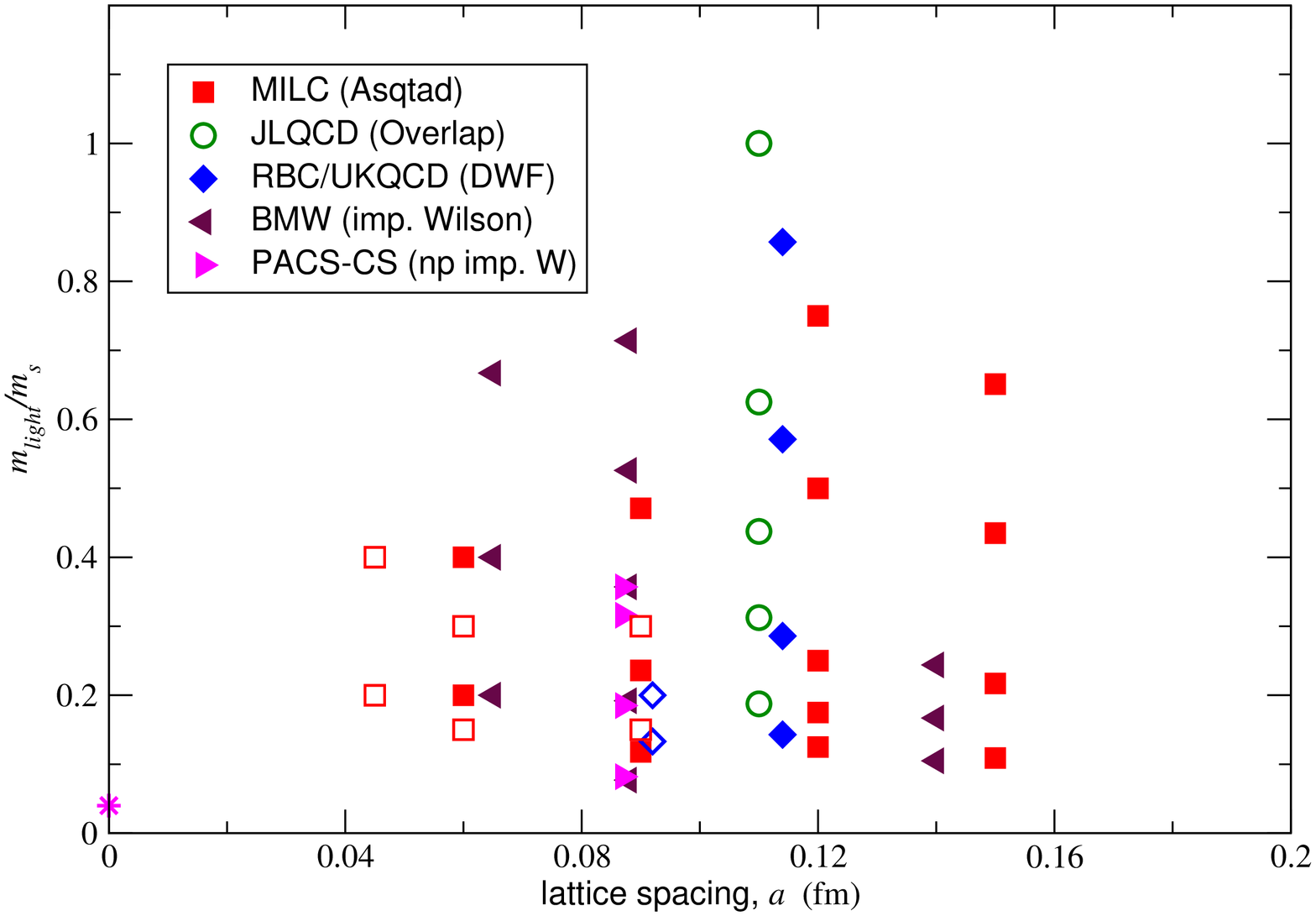}
\caption{Simulation parameters for ensembles with $n_f = 2+1$ showing
$m_l/m_s$ vs.~lattice spacing $a$. Filled symbols denote existing
ensembles. Unfilled symbols denote ensembles which are currently being
generated or planned. Red squares: MILC \cite{milc},
blue diamonds: RBC/UKQCD \cite{rbc}, purple left triangles:
BMW (improved Wilson) \cite{bmw}, pink right triangles: PACS-CS
(nonperturbatively improved Wilson) \cite{pacs-cs},
green circles: JLQCD (Overlap) \cite{jlqcd}.
The physical point is at $m_l/m_s = 1/25$
(pink burst). } \label{mvacomp}
\end{figure}

With the generation of the MILC ensembles, the stakes for
lattice QCD calculations have risen. We are now able to
calculate the simplest quantities to a few percent accuracy.
As always, repetition is desirable to test different lattice
methods against each other. To date, all lattice calculations
that include realistic sea quark effects use the MILC ensembles
with rooted Asqtad sea quarks. As mentioned in section~\ref{lqm}, the
Asqtad action carries the smallest computational cost of any
light quark action. Nevertheless, recently other collaborations have
started to generate ensembles with different sea quark actions.
An overview is given in Figure~\ref{mvacomp}.
It shows that the other ensembles are
being generated at similar values of lattice spacing and light
quark masses as the MILC ensembles. The MILC collaboration
continues to generate new ensembles at even smaller lattice spacings.
They are also generating additional configurations for the existing
ensembles to further reduce statistical errors. As in experiment, in
lattice QCD smaller statistical errors give better control over
systematic errors. Hence, lattice calculations based on the MILC
ensembles will continue to become more accurate.

Any modern lattice QCD calculation that claims phenomenological
relevance must include a serious systematic error analysis. To be
relevant, it must include the correct number of sea quarks
(which all the ensembles shown in Figure~\ref{mvacomp} do). While
the masses of the light sea quarks are still unphysically large,
it must also include a study of the light quark mass dependence with
sufficiently small sea quark masses. Among other sources of error,
discretisation effects must be estimated and tested by repeating the
calculation at more than one lattice spacing.

In summary, we should expect to see lattice results from
these new ensembles in the near future. They will provide
important consistency tests of the lattice methods, and in particular
of the square root trick used by the MILC collaboration to generate
their $n_f = 2+1$ ensembles.

\begin{acknowledgments}
I thank Peter Boyle, Eduardo Follana, Benjamin Haas, Shoji Hashimoto,
Andreas Kronfeld, Laurent Lellouch, Peter Lepage, Shigemi Ohta,
Victor Pavlunin, Jim Simone,
Naoya Ukita, Doug Toussaint, and Ruth Van de Water for help in
preparing my talk. I thank the organizers for inviting me to this
very lively and interesting meeting. This work was supported
in part by the DOE under grant no.~DE-FG02-91ER40677.
\end{acknowledgments}

\bigskip 

\end{document}